\documentclass[lettersize,journal]{IEEEtran}
\usepackage{amsmath,amsfonts}
\usepackage{amssymb}
\usepackage{xcolor}        
\usepackage{soul}         
\soulregister\cite7
\soulregister\ref7
\usepackage{algorithmic}
\usepackage{algorithm}
\usepackage{array}
\usepackage{textcomp}
\usepackage{stfloats}
\usepackage{url}
\usepackage{verbatim}
\usepackage[caption=false,font=footnotesize]{subfig} 
\usepackage{graphicx}
\usepackage{cite}
\usepackage{lineno,hyperref}
\usepackage{tabularx}
\usepackage{multirow}
\usepackage{booktabs}
\usepackage{bbding}
\newcolumntype{C}{>{\centering\arraybackslash}X}
\begin{document}

\title{Efficient Asynchronous Federated Evaluation with Strategy \\Similarity Awareness for Intent-Based Networking \\in Industrial Internet of Things}

\author{Shaowen Qin, 
        Jianfeng Zeng, 
        Haodong Guo, 
        Xiaohuan Li, ~\IEEEmembership{Member,~IEEE,}\\
        Jiawen Kang, ~\IEEEmembership{Senior Member,~IEEE,}
        and Qian Chen
    \thanks{This work was supported in part by the National Natural Science Foundation of China under Grant U22A2054. ({\itshape Corresponding author: Xiaohuan Li}.)}
    \thanks{Shaowen Qin, Jianfeng Zeng and Haodong Guo are with the Guangxi University Key Laboratory of Intelligent Networking and Scenario System (School of Information and Communication, Guilin University of Electronic Technology), Guilin 541004, China (e-mails: qinsw@mails.guet.edu.cn; zengjianfengxx@163.com; 24022303026@mails.guet.edu.cn).}
    \thanks{Xiaohuan Li is with the Guangxi University Key Laboratory of Intelligent Networking and Scenario System (School of Information and Communication, Guilin University of Electronic Technology), Guilin 541004, China, and also with National Engineering Laboratory for Comprehensive Transportation Big Data Application Technology (Guangxi), Nanning 530001, China (e-mails: lxhguet@guet.edu.cn).}
    \thanks{Jiawen Kang is with the School of Automation, Guangdong University of Technology, Guangzhou 510006, China (e-mail: kavinkang@gdut.edu.cn).}
    \thanks{Qian Chen is with the School of Architecture and Transportation Engineering, GUET, Guilin, 541004, China (e-mail: chenqian@mails.guet.edu.cn).}
}  

\markboth{} 
{Shell \MakeLowercase{\textit{et al.}}: A Sample Article Using IEEEtran.cls for IEEE Journals} 


\maketitle

\begin{abstract}
 Intent-Based Networking (IBN) offers a promising paradigm for intelligent and automated network control in Industrial Internet of Things (IIoT) environments by translating high-level user intents into executable network strategies. However, frequent strategy deployment and rollback are impractical due to tightly coupled workflows and high downtime costs, while node heterogeneity and privacy constraints further complicate centralized strategy evaluation. To address these challenges, we propose a Federated Evaluation Enhanced Intent-Based Networking framework (FEIBN), which leverages large language models (LLMs) to translate user intents into structured strategy tuples and employs federated learning to support distributed strategy evaluation. To improve training efficiency and reduce communication overhead, we design a Strategy Similarity Aware Federated Learning mechanism (SSAFL), which selects nodes relevant to the task based on strategy similarity and resource status, and triggers asynchronous model uploads only when local updates are significant. Experiments demonstrate that the proposed method improves model accuracy, accelerates convergence, and reduces communication cost compared with the baselines.
\end{abstract}


\section{Introduction}
\label{SecI}
\IEEEPARstart{T}{he} Industrial Internet of Things (IIoT) has evolved substantially in both scale and complexity, becoming a core enabling technology for modern industrial systems \cite{r1, r2}. Intent-Based Networking (IBN) provides a promising paradigm for intelligent operation in IIoT by translating high-level user intents into executable network strategies \cite{r3, r4}. Unlike conventional network management intents that mainly involve routing or configuration updates, IIoT intents often involve task execution goals, device coordination rules, safety constraints, and temporal requirements \cite{r46}. For example, in a sensing-driven environment equipped with temperature, humidity, water-level, and ultrasonic modules, an engineer may express intents such as “increase the sampling priority of the ultrasonic sensing module” or “allocate more processing resources to the water-level monitoring zone.” To ensure safe and efficient system operation, such high-level instructions must be correctly interpreted and mapped to actionable IIoT strategies \cite{r5, r37}. Traditional intent analysis methods usually rely on predefined rules or simple semantic matching \cite{r47,r33,r34}, which limits their ability to handle diverse and context-dependent industrial intents. In contrast, Large Language Models (LLMs) \cite{r6} provide stronger semantic understanding capabilities and can support more accurate intent interpretation in IBN systems \cite{r43}.

However, accurate intent recognition alone is insufficient to ensure reliable strategy deployment. In IIoT systems, generated strategies may directly affect production workflows and safety-critical devices, where inappropriate deployment can lead to costly downtime \cite{r35,r36}. Therefore, candidate strategies need to be evaluated before deployment to reduce operational risks and avoid unnecessary interruptions \cite{r4}. Existing AI-based evaluation methods usually require operational and environmental data from multiple devices to be collected at a centralized server for model training and performance prediction. Nevertheless, IIoT nodes are typically distributed and heterogeneous, and their local data often contain sensitive information such as device parameters and operational states \cite{r7}, making centralized strategy evaluation difficult. Federated Learning (FL) \cite{r8,r9} provides a distributed collaborative learning paradigm that can support the training of strategy-evaluation models across IIoT nodes without exposing raw data \cite{r45}. In synchronous FL, the server must wait for all selected clients to upload their updates, causing faster clients to remain idle until the slowest ones finish. This straggler effect slows down training, reduces resource utilization, and delays model convergence \cite{r24,r25}. Asynchronous FL alleviates this problem by allowing the server to update the global model once client updates arrive \cite{r23}. Despite these advances, IBN in IIoT still faces the following challenges.
\begin{itemize}
\item[i.] In IIoT, IBN requires not only intent understanding and strategy generation, but also pre-deployment evaluation and feedback support. Without a unified framework that connects these procedures, it is difficult to ensure that generated strategies can be reliably evaluated before deployment and effectively operated in IIoT systems.


\item[ii.] Different strategies usually correspond to different execution conditions, action sets, and affected resources \cite{r38}, which means that not all nodes are equally useful for the current evaluation task. Since IIoT nodes differ in local data, historical strategy experience, and resource availability, it is difficult to identify nodes that can provide high value updates for federated strategy evaluation.

\item[iii.] In dynamic IIoT environments, strategy evaluation needs to respond to changing industrial conditions in time. However, frequent uploads of minor local updates may introduce redundant communication, while overly delayed uploads may weaken the timeliness of evaluation \cite{r10}. This makes it difficult to balance response efficiency and communication cost during federated strategy evaluation.
\end{itemize}

To address the above challenges, this paper develops an intent-driven federated evaluation approach for IBN in IIoT. The proposed approach integrates multimodal intent understanding, LLM-based strategy translation, distributed strategy evaluation, and adaptive asynchronous model updates into a unified workflow. By considering strategy relevance, node resource availability, and update significance, it aims to improve the reliability of pre-deployment strategy evaluation while reducing unnecessary communication overhead. The main contributions of this paper are listed as follows.
\begin{table*}[!b]
\centering
\caption{Summary of Related Work on IBN.}
\label{SummaryIBN}
\begin{tabularx}{\textwidth}{m{0.5cm}|p{3.5cm}|p{4.2cm}|X}
\toprule
\textbf{Ref.}  & \textbf{Focus} &\textbf{Insight} & \textbf{Advantages \& Limitations} \\
\midrule
 \cite{r15}&  End-to-end intent life cycle design including intent parsing, strategy generation, and closed-loop execution & Establishes a complete AI-driven IBN pipeline that transforms high-level intents into enforceable network strategie's through multi-stage processing &
\hangindent=1.2em $\checkmark$ Provides structured IBN architecture, covers full strategy workflow  
\par
\hangindent=1.2em $\times$ Lacks LLM-based semantic reasoning, limited validation under dynamic IIoT or heterogeneous environments
\\ \hline

\cite{r16}  & LLM-based natural-language intent extraction, entity recognition, and slot filling & Demonstrates that LLMs significantly improve intent interpretation accuracy in 5G core networks and reduce configuration ambiguity & \hangindent=1.2em $\checkmark$ Enhances understanding of telecom intents, improves mapping precision
\par
\hangindent=1.2em $\times$ Focus solely on extraction without supporting strategy evaluation, assurance, or runtime validation
\\ \hline

\cite{r17} & Runtime assurance, semantic drift detection, state-to-intent consistency verification & Introduces the concept of intent drift, enabling LLMs to detect mismatches between desired intents and actual network behaviors & 
\hangindent=1.2em $\checkmark$ Strong in assurance and runtime monitoring, provides a new conceptual model
\par
\hangindent=1.2em $\times$ No intent translation or strategy generation; performance relies heavily on drift model robustness
\\ \hline

\cite{r18} & Agentic AI–based intent decomposition, multi-agent orchestration, and tool-enabled execution & Proposes an agentic intent-processing pipeline that decomposes industrial intents into actionable tasks via LLM-based multi-agent collaboration & 
\hangindent=1.2em $\checkmark$ Strong alignment with Industry 5.0, enables autonomous planning and execution
\par
\hangindent=1.2em $\times$ Conceptual and lacks network-level strategy evaluation, not tailored for communication constraints or heterogeneity
\\ 
\bottomrule
\end{tabularx}
\end{table*}

\begin{table*}[!t]
\centering
\caption{Summary of Related Work on FL.}
\label{SummaryFL}
\begin{tabularx}{\textwidth}{m{0.5cm}|p{3.5cm}|p{4.2cm}|X}
\toprule
\textbf{Ref.}  & \textbf{Focus} &\textbf{Insight} & \textbf{Advantages \& Limitations} \\
\midrule
\cite{r10}&  Asynchronous aggregation under heterogeneous device states & Improves model freshness by adjusting aggregation timing according to client states &
\hangindent=1.2em $\checkmark$ Better stability under asynchronous updates  
\par
\hangindent=1.2em $\times$ Does not distinguish task relevance among clients and lacks mechanisms to prevent low-value or irrelevant updates from harming global convergence
\\ \hline

\cite{r11} & Similarity-aware personalized FL & Uses confidence estimation and similarity weighting to improve personalized performance & 
\hangindent=1.2em $\checkmark$ Higher accuracy for heterogeneous autonomous devices.
\par
\hangindent=1.2em $\times$ Does not address client participation strategy and overlooks the impact of unreliable or inconsistent updates on training efficiency
\\ \hline

\cite{r12} & Task-grained knowledge sharing for heterogeneous task sequences & Shares compact task knowledge to support continual learning across diverse edge tasks & 
\hangindent=1.2em $\checkmark$ Strong support for heterogeneous tasks with reduced communication cost
\par
\hangindent=1.2em $\times$ Does not handle asynchronous participation and lacks a mechanism to prioritize high-value contributors under dynamic edge conditions
\\ \hline

\cite{r13} & Client clustering and personalized lightweight patches & Forms intrinsic client groups to improve personalization under non-IID data & 
\hangindent=1.2em $\checkmark$ Strong personalization capability
\par
\hangindent=1.2em $\times$ elies on fixed cluster structures and lacks adaptive handling of dynamic client states
\\
\bottomrule
\end{tabularx}
\end{table*}


\begin{itemize}
\item We propose a Federated Evaluation Enhanced IBN framework (FEIBN) that combines multimodal alignment with LLMs to translate user intents into structured strategy tuples and integrates federated learning to support distributed strategy evaluation. FEIBN enables candidate strategies to be evaluated before deployment, thereby supporting safer and more reliable operation in IIoT.

\item We design a Strategy Similarity Aware Federated Learning mechanism (SSAFL) for distributed strategy evaluation. SSAFL prioritizes nodes based on strategy relevance and resource availability, allowing nodes more suitable for the current evaluation task to participate in federated training, thereby improving training efficiency and accelerating model convergence.

\item We develop an adaptive asynchronous update mechanism to reduce redundant communication during federated strategy evaluation. By setting similarity-aware upload thresholds, the mechanism allows nodes to upload local model updates only when their update magnitudes are significant. It reduces unnecessary model transmissions while maintaining the effectiveness of asynchronous aggregation.
\end{itemize}

The rest of the paper is organized as follows. Section~\ref{SecII} reviews the related work. Section~\ref{SecIII} presents the system model of the proposed FEIBN framework. Section~\ref{SecIV} details the design of the SSAFL. Section~\ref{SecV} presents the experimental setup and results. Section~\ref{SecVI} concludes the paper.

\section{Related Work}
\label{SecII}
\subsection{Intent-Based Networking}
IBN abstracts user requirements into high-level intents and automatically maps them to executable network strategies, offering a promising approach for achieving automated and intelligent network control in IIoT environments. With the advancement of artificial intelligence, some studies have leveraged AI-driven methods to enhance intent understanding. The authors in 
\cite{r15} introduced an AI-powered IBN architecture that automates the mapping from user intents to strategy execution logic. In addition, LLMs have also been explored as powerful tools for semantic alignment in IBN systems. The authors in \cite{r16} designed a custom LLM-driven framework for extracting intents in 5G core networks, showcasing significantly improved intent interpretation for strategy generation. The authors in \cite{r17} proposed an LLM-guided assurance mechanism to detect and correct intent drift in real time, ensuring strategy consistency. The authors in \cite{r18} introduced an industrial Agentic AI system that decomposes high-level intent into executable control flows using LLM agents, demonstrating feasibility in predictive maintenance scenarios. A summary of related studies is provided in Table \ref{SummaryIBN}.

However, due to the involvement of multiple production-line devices in IIoT environments, it is impractical to frequently deploy and roll back strategies in real-world industrial operations. IBN in IIoT still lacks effective mechanisms for verifying the effectiveness of strategies prior to deployment. In addition, the heterogeneity and distributed nature of IIoT nodes further exacerbate the complexity of centralized strategy evaluation and coordination.

\subsection{Federated Learning}
Due to the wide distribution of IIoT nodes and the high sensitivity of local data, federated learning often faces practical challenges such as task diversity, heterogeneous device capabilities, and varying strategy applicability across clients, making it difficult to meet the personalized and efficient requirements of IBN strategy evaluation. To address this, several studies have focused on task-aware federated learning approaches. The authors in \cite{r11} proposed a federated learning method that emphasizes task similarity among clients by adopting a confidence-aware weighted aggregation strategy, guiding clients with similar tasks to share model parameters more closely and thus improving knowledge transfer efficiency. The authors in \cite{r12} introduced a task-granular knowledge aggregation method, where each client selectively integrates only the task-relevant parts of global knowledge to reduce communication costs and mitigate catastrophic forgetting. The authors in \cite{r13} presented a personalized federated learning framework based on task similarity, which dynamically adjusts aggregation weights to enhance collaborative effectiveness across tasks. The authors in \cite{r10} developed an asynchronous federated learning framework tailored for heterogeneous IoT environments, utilizing asynchronous updates and adaptive aggregation to improve training efficiency and overall stability under non-synchronous conditions. A summary of related studies is provided in Table \ref{SummaryFL}.

\begin{figure*}[t]
	\centering
		\includegraphics[width=0.8\textwidth]{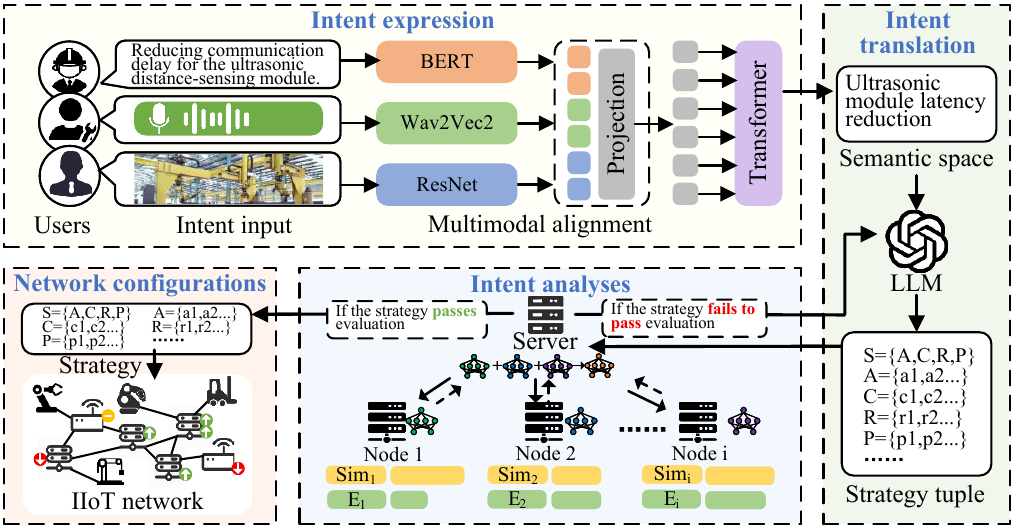}
	\caption{The FEIBN framework for IIoT. The framework supports intent-driven strategy generation and verification across distributed IIoT nodes. Within this framework, an LLM translates multimodal user intents into strategy tuples, and asynchronous federated strategy evaluation is performed based on a similarity-aware node selection mechanism.}
	\label{fig1}
\end{figure*}

However, most of these methods are designed for general purpose federated learning tasks and are not directly tailored to strategy evaluation in IBN. In particular, they seldom consider the relevance between the current strategy and each node's historical strategies, making it difficult to select nodes that can provide high value updates for the current evaluation task. Therefore, in IIoT scenarios with heterogeneous devices and dynamically changing node states, existing approaches still struggle to balance training efficiency, communication cost, and evaluation reliability.

\section{Federated Evaluation Enhanced Intent-Based Networking with LLM}
\label{SecIII}
To enable intelligent intent understanding and distributed strategy evaluation in IIoT environments, we propose the FEIBN, as illustrated in Fig. \ref{fig1}. The FEIBN framework consists of four core modules: intent expression, intent translation, intent analyses, and network configuration. First, in the intent expression module, users express their intents in multiple modalities, which are processed by a multimodal alignment module composed of pretrained encoders to extract semantic features. These features are then fused and interpreted in the intent translation module by an LLM, producing a structured strategy tuple. Next, in the intent analysis module, strategy validation is initiated across distributed IIoT nodes. A similarity-aware participation scoring mechanism evaluates each node’s relevance to the current strategy and its available resources. Based on this score, a subset of high-quality nodes is selected to participate in local training. Each participating node computes the magnitude of its local model update and uploads the update only if it exceeds a dynamic threshold, ensuring communication efficiency. Finally, in network configuration module, the central server aggregates these updates to evaluate the strategy effectiveness, and, if validated, the strategy is deployed to the industrial control system for execution. The main notations used in this paper are shown in Table \ref{tab:notation}.

\begin{table}[t]
\caption{SUMMARY OF MAIN NOTATIONS}
\label{tab:notation}
\centering
\begin{tabular}{lp{0.65\linewidth}}
\hline
\textbf{Notation} & \textbf{Description} \\
\hline
$a_g$ & Scaling factor controlling sensitivity to threshold differences in condition similarity. \\
$b_m$ & Bias parameter in the projection for modality $m$. \\
$d_i$ & Magnitude of local model update at client $i$. \\
$f(\theta;x_j,y_j)$ & Loss function on sample $(x_j,y_j)$, typically a regression loss such as MSE. \\
$g_n$ & A goal element, defined by a metric, relational operator, and threshold. \\
$h(g,g')$ & Pairwise condition similarity between two goals. \\
$m$ & Modality index, such as text, audio, or vision. \\
$q(t)$ & Set of client indices whose updates arrive within event window $t$. \\
$u_i$ & Normalized CPU utilization of node $i$. \\
$w$ & aggregation weight. \\
$z^m$ & Projected embedding of modality $m$. \\

$A_{i,j}$ & Action set of the $j$-th historical strategy of node $i$. \\
$B_i$ & Normalized available bandwidth of node $i$. \\
$C_{i,j}$ & Condition set of the $j$-th historical strategy of node $i$. \\
$D_i$ & Local dataset of node $i$. \\
$E'$ & Entities or resources in the executable strategy tuple. \\
$F_i(\theta)$ & Local objective function at node $i$. \\
$G'$ & Executable goal set in the strategy. \\
$H_i$ & Suitability score of node $i$, combining similarity and resources. \\
$I$ & Total number of IIoT nodes. \\
$S$ & Structured intent tuple including user, goals, entities, actions, and time. \\
$Sim_i$ & Strategy similarity of node $i$. \\
$T_i$ & Number of local rounds completed by client $i$. \\
$U$ & User in the intent tuple. \\
$\Gamma_i^t$ & Communication cost of client $i$ at round $t$. \\
$\beta_1,\beta_2$ & Weights for similarity and resource in the suitability score. \\
$\delta_1,\delta_2$ & Weights for CPU and bandwidth in the resource score. \\
$\epsilon_i$ & Client-specific upload threshold. \\
$\eta$ & Local learning rate used in client training. \\
$\lambda_s$ & Scaling factor in adaptive threshold design. \\
$\mu_g,\mu_{g'}$ & Threshold values of conditions in goals $g$ and $g'$. \\
$\nu$ & Convergence tolerance in global objective. \\
$\theta_i^t$ & Local model parameter of node $i$ at time $t$. \\
$\gamma_1,\gamma_2,\gamma_3$ & Weights for action, condition, and resource similarity components. \\
$\tau_s$ & Threshold for selecting clients based on suitability score. \\
$y,\hat{y}$ & True and predicted outputs in regression-based evaluation. \\
\hline
\end{tabular}
\end{table}

\subsection{Intent Expression}
In IIoT environments, user intents may appear in diverse forms. For instance, a field operator may issue a voice command such as \textit{“prioritize safety strategies in the pump station due to abnormal vibration,”} a supervisor may send a text message like \textit{“increase throughput of line B by 10\% within 2 hours,”} while a monitoring system may provide a visual signal indicating machine overheating. These heterogeneous inputs contain complementary cues, text captures explicit goals, audio conveys urgency or priority, and vision reflects real-time physical states. 

We develop an intent expression module in FEIBN that projects text, audio, and images into a unified semantic space, ensuring that intents expressed across diverse industrial contexts can be uniformly interpreted and effectively processed. To achieve consistent interpretation, these heterogeneous signals are first encoded into modality-specific embeddings. Specifically, textual sequences are processed using a pretrained BERT encoder \cite{r26}, audio waveforms are transformed into latent representations by Wav2Vec2 \cite{r27}, and visual inputs are converted into high-level semantic features via ResNet \cite{r28}. These models are selected for their strong generalization ability and proven robustness across multiple tasks, making them suitable for industrial scenarios where signals exhibit diverse formats and noise patterns. Since these encoders produce representations in different spaces, a learnable linear projection is applied to map each modality into a unified latent space as follows:
\begin{equation}
\begin{split}
z_m=W_m h_m+b_m\label{eq17} 
\end{split}
\end{equation}
where $m$ represents the type of input. $W_m$ and $b_m$ are trainable parameters. The projected embeddings from multiple modalities are concatenated and then processed by a Transformer encoder, which models cross-modal dependencies and contextual relations among modalities. For example, it can associate the spoken phrase “slow down” with a corresponding visual cue of increasing conveyor-belt speed, thereby reinforcing semantic coherence. Through self-attention, the Transformer learns which modality carries dominant information for a given intent. The resulting fused representation $z$ serves as a comprehensive semantic descriptor that combines textual precision, auditory intent strength, and visual situational awareness. Finally, the fused representation $z$ is passed to an LLM (e.g., GPT \cite{r29}, DeepSeek \cite{r31}, and LLaMA \cite{r32}), providing a coherent semantic interface for LLM-based intent translation. This unified representation enables the LLM to reason over structurally consistent inputs, thereby improving the accuracy and stability of strategy generation and forming the foundation for subsequent strategy generation and strategy-similarity evaluation.

\subsection{Intent Translation}
To ensure that high-level intents can be accurately and efficiently deployed in IIoT networks, the unified semantic representation needs to be converted into executable network strategies. In the intent translation module, the LLM is used for strategy generation, transforming abstract multimodal semantics into actionable and verifiable network configurations. The output strategy generated by an LLM is formally represented as a structured intent tuple, denoted as
\begin{equation}
\begin{split}
S=<U, G, E, A, T>\label{eq1} 
\end{split}
\end{equation}
where $U$ denotes the user who defines the intent. $G$ denotes the objective. $E$ denotes the infrastructure for deploying the intent. $A$ denotes the set of actions to be executed in the network. $T$ denotes the period that the required service is scheduled to occur.

Once the intent tuple S is received, the Central Strategy Engine transforms it into executable strategy tuples, denoted as
\begin{equation}
\begin{split}
S^\prime=<U,G^\prime,E^\prime,A^\prime,T>\label{eq2} 
\end{split}
\end{equation}
where $G^\prime=<g_1,g_2,\ldots,g_n>$ denotes the set of goals, representing the target objectives that the strategy aims to achieve, where each goal $g_n$ can be formally expressed as $g_n=\ell_n\vartriangleright\theta_n$, with $\ell_n$ representing a metric, $\theta_n$ a threshold, and $\vartriangleright$ a relational operator (e.g., $>,<,\geq,\le$). $E^\prime=<e_1,e_2,\ldots,e_k>$ identifies the devices or resources affected by the strategy. $A^\prime=<a_1, a_2, \ldots, a_k>$ denotes the set of actions to be executed, with each action $a_k$ indicating a concrete operational step. $T$ denotes the period in which the strategy is expected to take effect. Specifies when the required service behavior should be enacted. Below, we provide an example output of the intent translation module for a user intent such as \textit{“reduce communication delay for the ultrasonic sensing module”}: $<’operator\_02’$, $\{’metric’:’latency’$, $’operator’:’<’$, $’value’:15\}$, $’ultrasonic\_module’$, $\{’type’:’QoS\_adjustment’$, $’params’:\{’priority’:5\}\}$, $(0, 600s)>$. The field U identifies the initiating user (operator$\_$02). The goal set $G^\prime$ specifies that the end-to-end latency should be kept below 15ms. The entity set $E^\prime$ indicates that the strategy targets the ultrasonic module. The action set $A^\prime$ describes a concrete network operation, namely a QoS adjustment that raises the scheduling priority of the corresponding traffic to level 5, encoded through the type and params fields. Finally, the time field $T$ defines a 600 second window during which this strategy should be enforced.

\subsection{Intent Analyses with LLM}
In IIoT environments, where production lines are fixed and downtime costs are high, it is impractical to validate strategies through frequent real-world deployments. Therefore, the intent analysis module is designed to evaluate the effectiveness of strategies in a distributed manner prior to actual deployment.

The intent analysis module initiates a federated learning based on strategy $S$ to collaboratively train a predictive model capable of evaluating the strategy. We represent the set of IIoT nodes involved as $I_{node}=\{1,\ldots,i,\ldots,I\}$. Each node $i\in I_{node}$ possesses a local dataset $D_i$, consisting of samples $\left(x_j,y_j\right)$, where $x_j$ denotes the input feature, and $y_j$ is the corresponding label indicating whether strategy $S$ is suitable under the local context. Let $w$ denote the shared model parameter and $f\left(w;x_j;y_j\right)$ be the loss function on the $j$-th sample. The local objective of node $i$ is defined as
\begin{equation}
\begin{split}
F_i\left(\theta\right)=\frac{1}{\left|D_i\right|}\sum_{\left(x_j,y_j\right)\in D_i} f\left(\theta;x_j;y_j\right)\label{eq3} 
\end{split}
\end{equation}
where $\left|D_i\right|$ represents the size of the dataset $D_i$. Therefore, the loss function $F\left(w\right)$ of the server side can be calculated as
\begin{equation}
\begin{split}
F\left(\theta\right)=\sum_{i=1}^{I}\frac{\left|D_i\right|F_i\left(\theta\right)}{\left|D\right|}\label{eq4} 
\end{split}
\end{equation}
where $\left|D\right|=\sum_{i=1}^{I}\left|D_i\right|$. According to the above loss function, the optimization objective of FL can be formulated as
\begin{equation}
\begin{split}
\theta^*=\arg\min_\theta F(\theta)\label{eq5} 
\end{split}
\end{equation}
where $w^*$ is the optimal global model.

After convergence, the global model outputs a deployability score for strategy $S$, reflecting the probability that $S$ can achieve its goal set $G'$ across heterogeneous IIoT nodes. High-scoring strategies are approved for configuration and deployment, while low-scoring ones are refined or re-evaluated. Furthermore, to achieve efficient and scalable federated evaluation across heterogeneous IIoT nodes, a strategy similarity aware federated learning mechanism is employed. which is discussed in Section~\ref{SecIV}.

\subsection{Network Configurations}
After the intent analysis module verifies that a candidate strategy $S$ satisfies the performance and safety requirements, the strategy proceeds to the network configuration stage for deployment in the industrial environment. In this stage, the verified intent is translated into executable control commands that are delivered to the corresponding network elements and industrial devices.

The action set $A'$ is mapped to concrete configuration commands for each entity $e\in E'$, which can be abstracted as
\begin{equation}
    c_e = \Phi_e(S'), \quad \forall e \in E',
\end{equation}
where $c_e$ denotes the configuration state of entity $e$ and $\Phi_e(\cdot)$ represents the configuration mapping implemented by the controller.

During deployment, real-time telemetry data, such as latency, bandwidth utilization, equipment status, and workload metrics, are continuously collected and compared with the expected performance objectives defined during strategy generation. Let $\tilde{\ell}_n(t)$ denote the measured value of metric $\ell_n$ at time $t$. The satisfaction indicator of goal $g_n$ at time $t$ is defined as
\begin{equation}
    \sigma_n(t) =
    \begin{cases}
        1, & \text{if } \tilde{\ell}_n(t)\ \triangleright\ \theta_n,\\[2pt]
        0, & \text{otherwise.}
    \end{cases}
\end{equation}
and the overall satisfaction of strategy $S'$ at time $t$ is given by
\begin{equation}
    J_S(t) = \prod_{n=1}^{|G'|} \sigma_n(t),
\end{equation}
where $J_S(t)=1$ indicates that all goals in $G'$ are satisfied and $J_S(t)=0$ otherwise.

Over a deployment window $T$, the empirical satisfaction probability of $S'$ is computed as
\begin{equation}
    p_S = \frac{1}{|T|} \sum_{t \in T} J_S(t),
\end{equation}
where $|T|$ denotes the number of observation instants in $T$. When deviations from the desired targets are detected, e.g., when $p_S < p_{\min}$ for a predefined reliability threshold $p_{\min} \in (0,1)$, the system dynamically adjusts configuration parameters or triggers re-verification through the federated evaluation process. This adaptive feedback ensures that each deployed strategy remains valid and stable even under varying network conditions or workload fluctuations.

The network configuration module bridges the gap between intent-level decision-making and operational execution. It ensures that every strategy applied in the IIoT system is validated, explainable, and adaptive to dynamic industrial environments, thereby enabling trustworthy and autonomous operation within the intent-based networking framework.

\section{Strategy Similarity Aware Asynchronous Federated Learning}
\label{SecIV}
\begin{figure}[!t]
\centering
\includegraphics[width=0.98\linewidth]{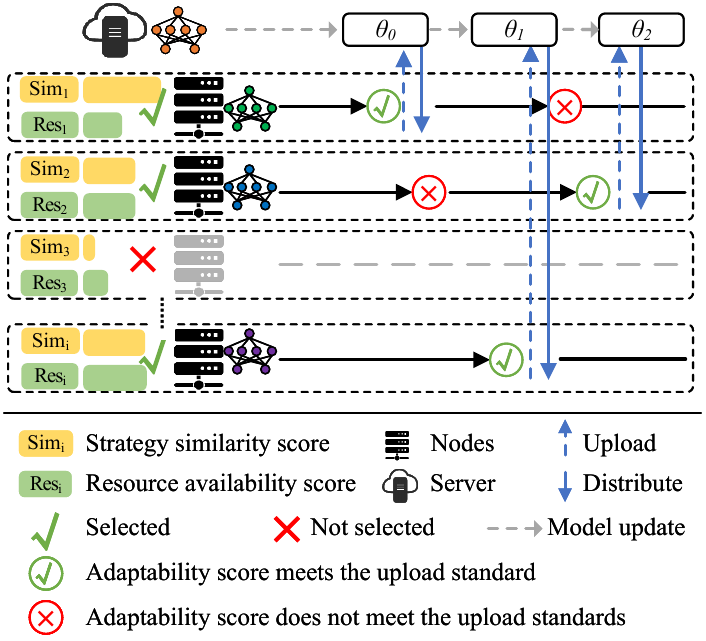}
\caption{The process of SSAFL. Nodes first evaluate their strategy similarity and resource availability, and only those satisfying the selection criteria join the current round. Selected nodes receive the global model, perform local training, and compute adaptability scores. Nodes whose adaptability meets the upload threshold send updates asynchronously to the server. The server aggregates the received updates and distributes the refined global model.}
\label{fig10}
\end{figure}

In IFEIBN, federated learning is employed to enable distributed strategy evaluation. Traditional FL methods are primarily designed for general-purpose tasks and therefore cannot effectively distinguish which nodes possess the historical knowledge most relevant to the current strategy, nor can they leverage such relevance to guide efficient model training. To address this limitation, we design SSAFL, which introduces a strategy similarity metric to quantify the semantic closeness between the current strategy and each node’s historical strategy set. SSAFL adaptively selects nodes that are both semantically aligned and resource-sufficient, ensuring that nodes with the highest contribution value participate more substantially in the FL process. Furthermore, SSAFL incorporates a similarity-driven asynchronous update mechanism to prioritize meaningful model uploads and aggregation. As shown in Fig. \ref{fig10}, each node evaluates its strategy similarity score and resource availability score, which together determine its adaptability score for participation in the current federated round. Nodes with adaptability scores exceeding the upload threshold are selected to upload their local model updates to the server, while the others are temporarily excluded from the aggregation process. The server then performs a weighted aggregation to update the global model and redistributes it to the nodes that contributed updates. This mechanism ensures that nodes with higher semantic relevance to the current strategy and sufficient computational resources contribute more effectively to the global optimization process.

\subsection{Strategy Similarity Based Node Selection Scheme}
To accurately quantify the similarity between the current strategy $S$ and the historical strategies maintained by nodes in FEIBN, we design a strategy similarity metric. This metric is decomposed into three components: action similarity, condition similarity, and resource similarity. The strategy similarity score of node $i$ for strategy $S$ is defined as
\begin{equation}
\begin{split}
Sim_i(S) = \gamma_1 \left| \frac{A \cap A_{i,j}}{A \cup A_{i,j}} \right| 
+ \gamma_2 \frac{1}{|\mathcal{G}|} \sum_{c \in \mathcal{C}} \max_{c' \in c_{i,j}} h(g, g')\label{eq6} 
\end{split}
\end{equation}
where $\gamma_1,\gamma_2\in\left[0,1\right]$ are weights satisfying $\gamma_1+\gamma_2=1$. $\frac{\left|A\cap A_{i,j}\right|}{\left|A\cup A_{i,j}\right|}$ denotes the action similarity, which evaluates the overlap between the action sets of the two strategies, and is calculated using the Jaccard similarity coefficient. $\left|\bullet\right|$ denotes the cardinality of a set, a value of 1 indicates identical action sets, and a value of 0 indicates no common actions. $\frac{1}{|\mathcal{G}|} \sum_{c \in \mathcal{C}} \max_{c' \in c_{i,j}} h(g, g')$ denotes the condition similarity, which measures the degree of alignment between the conditions under which actions are applied. $h\left(g,g^\prime\right)$ is a pairwise condition similarity function, which can be defined as
\begin{equation}
\begin{split}
h(g, g') = 
\begin{cases}
\exp\left( -a_g \left( \frac{|\mu_g - \mu_{g'}|}{\mu_g} \right) \right), & \text{if metrics match} \\
0, & \text{otherwise}
\end{cases}\label{eq7} 
\end{split}
\end{equation}
where $\mu_g$ and $\mu_{g^\prime}$ are the thresholds of conditions $c$ and $c^\prime$. $\alpha_g>0$ is a scaling factor controlling the sensitivity to threshold differences. $h\left(g,g^\prime\right)$ adopts an exponential decay formulation to measure the semantic closeness between two intent conditions. It ensures that two conditions exhibit a high similarity score when they involve the same performance metric and their thresholds are close, while their similarity decreases rapidly as the threshold gap widens \cite{r39}. Such behavior naturally reflects the semantics of intent conditions in IBN, where even small deviations in latency, loss, or throughput constraints may lead to significantly different operational requirements.

To efficiently select IIoT nodes for federated training in FEIBN, we design a suitability score $H_i$ that evaluates each node's potential contribution based on two key factors: strategy similarity and resource availability. The suitability score guides the asynchronous training process by preferentially selecting nodes most relevant to the current validation task. For a node $i$ and a target strategy $S$, the suitability score $H_i$ is defined as
\begin{equation}
\begin{split}
H_i=\beta_1{Sim}_i\left(S\right)+\beta_2{Res}_i\label{eq8} 
\end{split}
\end{equation}
where ${Res}_i$ denotes the current resource status of the node $i$. $\beta_1,\beta_2\in\left[0,1\right]$ are weights satisfying $\beta_1+\beta_2=1$.

The resource availability score ${Res}_i$ captures the computational and communication readiness of the node and is computed as
\begin{equation}
\begin{split}
{Res}_i=\delta_1\left(1-U_i\right)+\delta_2B_i\label{eq9} 
\end{split}
\end{equation}
where $U_i$ denotes the normalized CPU utilization of node $i$. $B_i$ denotes the normalized available communication bandwidth. $\delta_1,\delta_2\in\left[0,1\right]$ are resource-specific importance weights satisfying $\delta_1+\ \delta_2=1$.

Given a threshold $\tau_s$, node $i$ is selected to participate in the current training round if $H_i\geq\tau_s$. Otherwise, it remains idle for this training. 

\subsection{Adaptive Model Training and Updating}

To efficiently validate strategies in FEIBN, we adopt an asynchronous FL approach, where node participation and model updates occur independently based on each node's readiness and relevance to the current validation task.
Upon receiving the current validation strategy $S$ from the server, each selected node $i$ initiates local training. Each node computes the $L_2$ norm of its local model update, denotes as
\begin{equation}
\begin{split}
\left\| \Delta \theta_i^t \right\|_2 = \left\| \theta_i^t - \theta_{\text{global}}^{t-1} \right\|\label{eq10} 
\end{split}
\end{equation}
where $\theta_i^t$ denotes the node’s local model parameters after training. $\theta_{global}^{t-1}$ denotes the latest global model parameters received by the node before local training. We define $\left\| \Delta \theta_i^t \right\|_2$ as the distance between the model trained by node $i$ and the global model.

We set an update threshold for the node to upload its update only when it exceeds this threshold. The update threshold is defined as
\begin{equation}
\begin{split}
\epsilon_i=\epsilon_{base}\times\left(1+\lambda_s\times\left(1-{Sim}_i\left(S\right)\right)\right)
\label{eq11} 
\end{split}
\end{equation}
where $\epsilon_{base}$ is the base threshold value. $\lambda_s$ is the scaling factor controlling the influence of similarity on the threshold.

The node uploads its model update $\Delta \theta_i^t$ to the server if and only if $\left\| \Delta \theta_i^t \right\|\geq\epsilon_i$. Otherwise, the node will continue to train its local model until the model distance reaches a threshold, thus avoiding unnecessary communication overhead.

When a node $i$ uploads its local model update $\Delta \theta_i^t$ to the server after passing the upload threshold, the server performs asynchronous aggregation immediately without waiting for other nodes. The server receives $\Delta \theta_i^t$ and computes the preliminary weight $w_i^\prime$ as
\begin{equation}
\begin{split}
w_i^\prime={Sim}_i\left(S\right)\times \left\|\Delta \theta_i^t\right\|_2
\label{eq12} 
\end{split}
\end{equation}

To avoid the situation where important nodes contribute insignificantly due to small update magnitudes, we introduce a minimum weight protection mechanism. The final aggregation weight $w_i$ is defined as
\begin{equation}
\begin{split}
w_i = \left( w_{\text{min}}, \frac{w_i'}{\sum_{j \in Q(t)} w_j'} \right)
\label{eq13} 
\end{split}
\end{equation}
where $w_i$ is a predefined minimum weight threshold. $Q\left(t\right)$ denotes the set of nodes whose updates have been received by the server in the current aggregation server. The server asynchronously updates the global model using:
\begin{equation}
\begin{split}
\theta_{\text{global}}^{t+1} = \theta_{\text{global}}^t + w_i \Delta \theta_i^t
\label{eq14} 
\end{split}
\end{equation}

\subsection{Problem Formulation}
We define the communication cost incurred by node $i$ after the $t$-th round of local training as $\Gamma_i^t$. If the local model $\theta_i^t$ satisfies $\left\| \Delta \theta_i^t \right\|\geq\epsilon_i$ and uploads the model, we define $\Gamma_i^t=1$. Therefore, the communication cost of the node is more formally expressed as
\begin{equation}
\begin{split}
\Gamma_i^t =
\begin{cases}
\Gamma_i^{t-1} + 1, & \text{if } \|\Delta \theta_i^t\|_2 \geq \epsilon_i \\
\Gamma_i^{t-1}, & \text{otherwise}.
\end{cases}
\label{eq15} 
\end{split}
\end{equation}

In the FEIBN, the objective of SSAFL is to minimize the overall communication cost throughout the federated validation process while ensuring that the final global model achieves acceptable validation accuracy. The communication cost of each client throughout the training process is abbreviated as $\Gamma_i=\sum_{t=1}^{T_i}\Gamma_i^t$, where $T_i$ denotes the number of rounds trained by the $i$-th node. Then, the objective function can be formulated as
\begin{equation}
\begin{split}
\begin{aligned}
&\min \sum_{i=1}^{I} \Gamma_i \\
&\text{s.t.} \quad F(\theta^t) \leq F(\theta^*) + \nu \\
&\qquad \|\Delta \theta_i^t\|_2 \geq \epsilon_i
\end{aligned}
\label{eq16} 
\end{split}
\end{equation}
where $\theta^\ast$ is the optimal FL training model, and $\nu$ is a constant. 

\subsection{Algorithm Design and Explanation}
\begin{algorithm}[t]
\caption{Client-side Local Training with Thresholded Upload}
\label{alg:ssafl-client}
\begin{algorithmic}[1]
\REQUIRE Received $(\theta^t,S,\epsilon_i)$; local data $D_i$; local epochs $E_i$; stepsize $\eta$
\STATE $\theta_i \leftarrow \theta^t$
\REPEAT
  \STATE // Local training against $F_i(w)$ in Eq. \eqref{eq3}
  \FOR{$e=1$ to $E_i$}
     \STATE $\theta_i \leftarrow \theta_i - \eta \nabla F_i(\theta_i)$
  \ENDFOR
  \STATE $\Delta\theta_i \leftarrow \theta_i - \theta^t$;\quad $d_i \leftarrow \|\Delta\theta_i\|_2$ \COMMENT{Eq. \eqref{eq11}}
  \IF{$d_i \ge \epsilon_i$}
     \STATE Upload $\Delta\theta_i$ (and metadata such as $|D_i|,t$) to server
     \STATE \textbf{wait} for next $\theta^{t+1}$ from server; \ $\theta^t\!\leftarrow\!\theta^{t+1}$;\ $\theta_i\!\leftarrow\!\theta^t$
  \ELSE
     \STATE Continue local training to accumulate updates
  \ENDIF
\UNTIL{receive signal from server}
\end{algorithmic}
\end{algorithm}

\begin{algorithm}[t]
\caption{Server-side SSAFL in FEIBN}
\label{alg:ssafl-server}
\begin{algorithmic}[1]
\REQUIRE Intent tuple $S=\langle U,G,E,A,T\rangle$, initial global model $\theta^0$;
weight sets $\gamma,\beta,\delta$; thresholds $\tau_s,\epsilon_{\text{base}}$; scale $\lambda_s$;
minimum weight $w_{\min}$; stopping tolerance $\nu$.
\STATE // Node scoring and selection uses Eqs. \eqref{eq6},\eqref{eq8},\eqref{eq9}
\FOR{each node $i$}
  \STATE Compute $Sim_i(S)$ by Eq. \eqref{eq6}; compute $Res_i$ by Eq. \eqref{eq9};
  \STATE $H_i \leftarrow \beta_1\,Sim_i(S)+\beta_2\,Res_i$ \COMMENT{Eq. \eqref{eq8}}
\ENDFOR
\STATE $P \leftarrow \{\,i \mid H_i \ge \tau_s\,\}$ \COMMENT{select by threshold}
\STATE // Personalized upload thresholds uses Eq. \eqref{eq11}
\FOR{each $i \in P$}
  \STATE $\epsilon_i \leftarrow \epsilon_{\text{base}}\big(1+\lambda_s(1-Sim_i(S))\big)$
  \STATE Send $(\theta^t,S,\epsilon_i)$ to client $i$
\ENDFOR
\STATE // Event-driven asynchronous aggregation uses Eqs. \eqref{eq10},\eqref{eq12},\eqref{eq14}
\STATE Initialize a short event window $\Delta$ and buffer $Q(t)\!=\!\varnothing$
\LOOP
  \STATE \textbf{Upon} receiving update $\Delta\theta_i$ from any $i\in P$:
  \IF{$\|\Delta\theta_i\|_2 \ge \epsilon_i$}
    \STATE $Q(t)\leftarrow Q(t)\cup\{i\}$
    \STATE $w'_i \leftarrow Sim_i(S)\cdot \|\Delta\theta_i\|_2$ \COMMENT{Eq. \eqref{eq12}}
  \ENDIF
  \IF{window $\Delta$ expires or $|Q(t)| \ge 1$} 
     \STATE Min-weight protection \& normalization:
     \STATE $\tilde w_i \leftarrow \frac{w'_i}{\sum_{j\in Q(t)} w'_j}$,\quad
            $w_i \leftarrow \max\{w_{\min},\tilde w_i\}$,\quad
            $w_i \leftarrow \frac{w_i}{\sum_{j\in Q(t)} w_j}$
     \STATE $\theta^{t+1} \leftarrow \theta^{t} + \sum_{j\in Q(t)} w_j\,\Delta\theta_j$ \COMMENT{asynchronous update.}
     \STATE Update communication counters $\Gamma_i$ by Eq. \eqref{eq15}; clear $Q(t)$
     \STATE $t\leftarrow t+1$
  \ENDIF
  \IF{$F(\theta^t) \le F(\theta^\ast)+\nu$ \textbf{ or } $t\ge T_{\max}$}
     \STATE \textbf{break}
  \ENDIF
\ENDLOOP
\STATE \textbf{return} $\theta^t$
\end{algorithmic}
\end{algorithm}

The proposed SSAFL training process consists of two components: a client-side training procedure (Algorithm~\ref{alg:ssafl-client}) and a server-side coordination mechanism (Algorithm~\ref{alg:ssafl-server}). The client module handles local training and decides whether to upload updates based on an update norm threshold. The server module computes similarity-aware participation scores to select relevant nodes and aggregates valid updates asynchronously.

Algorithm~\ref{alg:ssafl-client} specifies the behavior of each participating client. 
After initialization with the received global model $\theta^t$, intent tuple $S$, and threshold $\epsilon_i$, the client performs local SGD training (Lines 3–6) according to Eq. \eqref{eq3}. 
It then computes the update $\Delta\theta_i = \theta_i - \theta^t$ and its L2 norm (Line 7, Eq. \eqref{eq10}). 
If the update magnitude exceeds the threshold $\epsilon_i$ (Lines 8–10), the client uploads $\Delta\theta_i$ to the server and waits for the next global model. 
Otherwise, it continues local training to accumulate larger updates (Lines 11–12), thereby avoiding unnecessary communication. 
The process repeats until a stop signal is issued by the server (Line 13). 

Algorithm~\ref{alg:ssafl-server} describes the federated training and aggregation procedure executed by the central server. 
Lines 1–4 compute the strategy similarity $Sim_i(S)$ (Eq. \eqref{eq6}) and resource availability $Res_i$ (Eq. \eqref{eq9}) for each node, then derive the suitability score $H_i$ using Eq. \eqref{eq8}. 
Line 5 selects nodes with $H_i \ge \tau_s$ to participate in training, ensuring only task-relevant and resource-capable nodes are involved. 
Lines 6–8 set personalized upload thresholds $\epsilon_i$ according to Eq. \eqref{eq11}, making high-similarity nodes more likely to upload. 
Lines 9–24 form the asynchronous event-driven loop: updates are received (Lines 11–13) and pre-weights $w'_i$ are computed (Eq. \eqref{eq12}); 
micro-batch aggregation is triggered (Lines 14–21), where minimum weight protection and normalization are applied before updating the global model via Eq.\eqref{eq14}. 
The communication counters $\Gamma_i$ are updated following Eq. \eqref{eq15}. 
Finally, convergence is checked (Lines 22–24) based on Eq. \eqref{eq16}, and the global model $\theta^t$ is returned (Line 26). 

The computational cost of SSAFL follows the same order as standard synchronous and asynchronous FL. For each aggregation event the server requires $O(|Q(t)|)$ operations to normalize weights and update the global model, where $|Q(t)|$ is the size of the micro-batch.  The client-side training follows the standard stochastic gradient descent (SGD) procedure and thus retains a complexity of $O(E_i\cdot|D_i|)$ per local epoch, where $E_i$ is the number of local epochs and $|D_i|$ the dataset size \cite{r40}.  Regarding communication, each client transmits its update only when the condition in Eq. \eqref{eq11} is satisfied. The expected number of transmissions per client is thereby reduced from $T_i$. 

According to the convergence conditions in the FL definition given by the literature \cite{r30}, the convergence of the proposed SSAFL update rule can be analyzed following the asynchronous federated optimization framework in \cite{r21}. The detailed convergence analysis of SSAFL is provided in Appendix~\ref{appendix}.

\section{Numerical Results}
\label{SecV}
\subsection{Experimental Setting}
We model the strategy validation problem as a regression task, where the goal is to predict the effectiveness score of a given strategy unit $S=<U, G, E, A, T>$ within its contextual environment. The predicted value $\hat{y}$ is employed to approximate the true deployable outcome $y$.

\textbf{Experimental Environment.} The experiments were carried out on a computing platform running Ubuntu 22.04.5 LTS, equipped with an Intel(R) Xeon(R) Platinum 8168 CPU @ 2.70GHz and 4 × NVIDIA RTX 3090 GPUs. The experiments were implemented in Python 3.9, with federated training simulated using the FedML framework.
\begin{figure}[!b]
\centering
\includegraphics[width=0.97\linewidth]{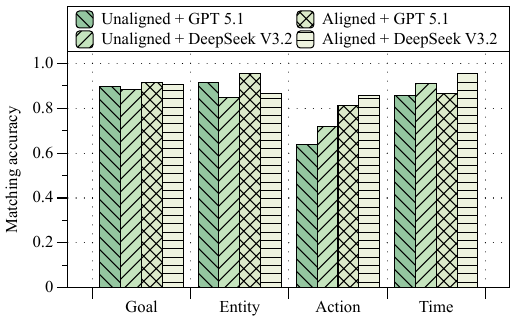}
\caption{Matching accuracy of strategy tuples under different LLMs.}
\label{fig5}
\end{figure}
\begin{figure}[!b]
\centering
\includegraphics[width=0.95\linewidth]{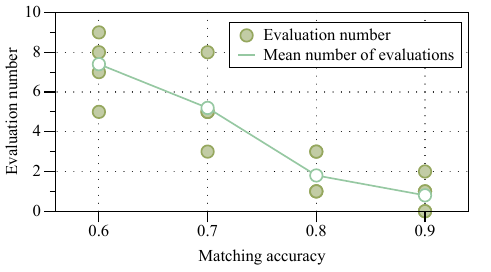}
\caption{Federated evaluation number under different matching accuracies.}
\label{fig6}
\end{figure}

\textbf{Datasets.} Datasets used in this experiment consists of two components. The first part is device parameter data obtained from the publicly available Edge-IIoTset \cite{r19} dataset, which includes real device operation logs and sensor parameters across various IIoT scenarios, thereby providing a representative reflection of IIoT node behaviors and characteristics under different operating conditions. The second part is intent-related data, which encompasses common business requirements in IIoT scenarios, such as bandwidth allocation, latency constraints, and energy–throughput trade-offs. In our setup, each client holds heterogeneous data sources, which naturally form a feature-skew non-IID distribution. Moreover, since the performance gains of SSAFL stem primarily from its similarity-aware scoring mechanism and asynchronous evaluation dynamics rather than from dataset-specific statistical properties, the same qualitative trends are expected to hold across different datasets.

\textbf{Methods.} We conducted comparative experiments on several federated learning strategies, including FedAvg \cite{r20}, Federated Asynchronous Learning (FedAsyn) \cite{r21}, and Semi-Asynchronous FL (SemiAsyn) \cite{r22}. In FedAsyn, the server updates the global model immediately upon receiving an update from any client, whereas in SemiAsyn, the server performs an update once it has received updates from top k clients.

\subsection{FEIBN Performance Comparison}
To evaluate the contribution of the multimodal alignment module, we analyze the accuracy of the generated strategy tuples $S$. As shown in Fig. \ref{fig5}, the alignment module notably improves the precision of slot prediction, with the most significant gain observed in the “Action”. This indicates that multimodal semantic fusion helps the model capture complex operational intents that cannot be fully expressed in text alone.

Fig. \ref{fig6} shows the variation in the number of federated evaluations under different matching accuracies. As alignment accuracy increases from 0.6 to 0.9, the number of evaluations performed decreases significantly. This result indicates that higher alignment quality enhances the semantic consistency of the strategies generated by the LLM (i.e., GPT-5.1 and DeepSeek-V3.2), enabling the system to make more accurate and confident decisions. Consequently, fewer redundant verifications are required, thereby improving the overall efficiency of the federated evaluation process.
\begin{figure}[!b]
\centering
\includegraphics[width=0.95\linewidth]{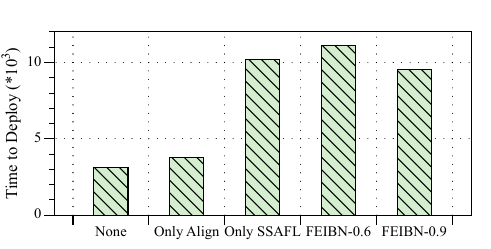}
\caption{Deployment time of different methods.}
\label{fig7}
\end{figure}

Fig. \ref{fig7} shows the total time required for strategy deployment across different methods. Adding only the alignment module slightly increases the deployment time due to the additional semantic parsing process. In contrast, FEIBN that integrates both alignment and federated evaluation results in a higher overall time cost, especially under lower alignment accuracy such as FEIBN-0.6, where more verification rounds are required. As the alignment accuracy increases to FEIBN-0.9, the deployment time decreases accordingly, indicating that improved alignment quality enhances the efficiency of federated validation and reduces the number of verifications.

\subsection{SSAFL Performance Comparison}
We randomly assign each node a subset of the training data from the dataset as its local training set, while the test set is retained on the server for performance evaluation. Following previous experimental settings, we compare SSAFL with other FL methods, with each method repeated five times. In addition, an ablation experiment is conducted on the adaptive model aggregation at the server side within SSAFL to verify the impact of this controllable factor on model training. When SSAFL does not include adaptive aggregation, it is denoted as SSAFL*. The experimental results are reported in Table \ref{tab1} as point estimates using the mean ± standard deviation.

\begin{table}[!t]
\centering
\caption{Comparison of different methods.}
\label{tab1}
\begin{tabularx}{\columnwidth}{CCCC}
\toprule
Method & MAE ↓ & RMSE ↓ & $R^2$ ↑ \\
\midrule
FedAvg   & 0.0637$\pm0.023$ & 0.0677$\pm0.035$ & 0.8398$\pm0.17$ \\
FedAsyn  & 0.0865$\pm0.036$ & 0.0921$\pm0.033$ & 0.7462$\pm0.28$ \\ 
SemiAsyn & 0.0594$\pm0.017$ & 0.0629$\pm0.022$ & 0.8840$\pm0.11$ \\
SSAFL*   & 0.0541$\pm0.023$ & 0.0597$\pm0.015$& 0.8703$\pm0.08$ \\
SSAFL    & \textbf{0.0497$\pm$0.011 } & \textbf{0.0521$\pm$0.017}& \textbf{0.9177$\pm$0.12} \\
\bottomrule
\end{tabularx}
\end{table}

\begin{figure}[!t]
\centering
\includegraphics[width=0.95\linewidth]{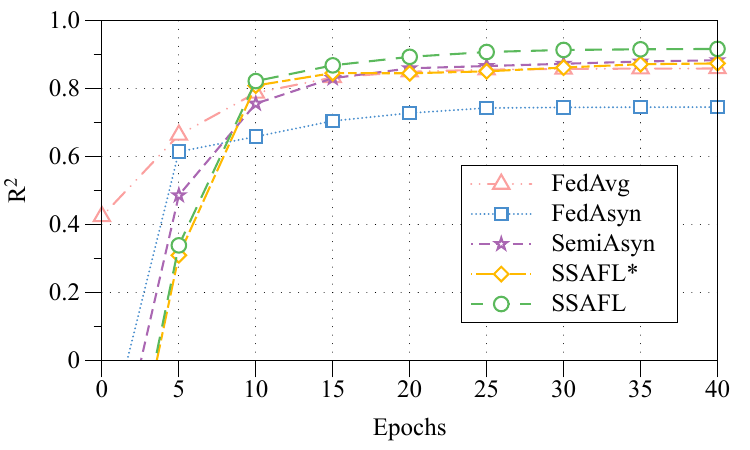}
\caption{Model training curves of different methods.}
\label{fig3}
\end{figure}

\begin{figure}[!t]
\centering
\includegraphics[width=0.95\linewidth]{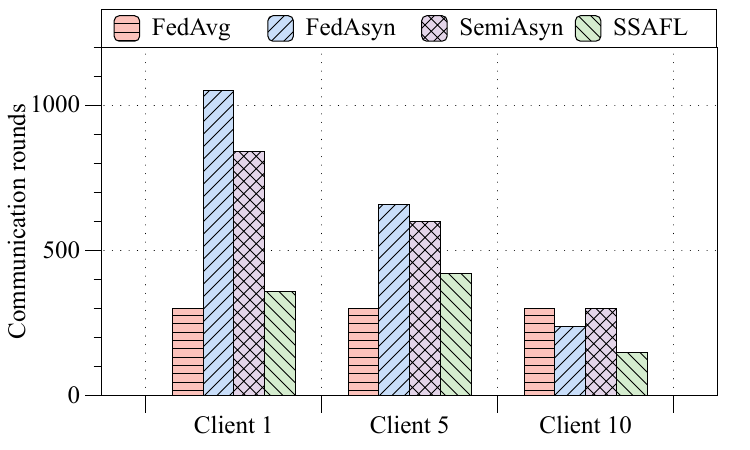}
\caption{Communication rounds of clients under different methods.}
\label{fig4}
\end{figure}

Fig. \ref{fig3} illustrates the R²-based training curves of five federated learning methods. SSAFL achieves the best training performance among all compared methods, converging to an R² of 0.89 within only 15 epochs. Its ablated variant SSAFL* also performs well, validating the effectiveness of similarity-aware node selection. FedAvg and FedAsyn show slower convergence and lower final R² scores, around 0.85 and 0.83 respectively. Overall, these results highlight the advantages of combining intent-aware participation scoring and asynchronous communication in federated strategy evaluation.

To evaluate the communication cost of different FL strategies under heterogeneous client latency, we configure Client 1, Client 5, and Client 10 as fast, medium, and slow clients, respectively, by assigning different local training times and upload delays. The experimental results are displayed in Fig. \ref{fig4}. Synchronous FedAvg produces identical communication rounds for all clients since each aggregation must wait for the slowest client. In contrast, asynchronous strategies show clear disparities. Fast clients upload much more frequently, while slow clients contribute fewer updates. SSAFL achieves the lowest communication rounds across all clients by suppressing redundant fast-client uploads and filtering low-impact updates from slow clients.

\section{Conclusion}
\label{SecVI}
In this paper, we have proposed FEIBN, a Federated Evaluation Enhanced Intent-Based Networking framework tailored for IIoT environments. FEIBN leverages large language models to align heterogeneous multimodal intents into structured strategy tuples, and integrates federated learning to achieve distributed strategy evaluation without exposing sensitive local data. To address the challenges of communication cost and training efficiency, we have further designed SSAFL, a Strategy Similarity Aware Federated Learning mechanism that combines similarity-aware node selection with adaptive asynchronous update thresholds.
The experiments have demonstrated that SSAFL significantly improves model accuracy and convergence speed while reducing communication overhead compared with existing synchronous and asynchronous baselines. The ablation studies further validated the effectiveness of similarity-aware participation scoring and adaptive aggregation in enhancing federated strategy evaluation.

{\appendix
\section*{Convergence Analysis of SSAFL}
\label{appendix}
According to the convergence conditions in the FL definition given by \cite{r30,r44}, it is assumed that Centralized Learning converges to the optimal model parameter $\theta^{(c)}$ and FL converges to the optimal model parameter $\theta^{(f)}$. If the gap between the two is small enough, that is, $\theta^{(f)}-\theta^{(c)}<\rho$ ($\rho$ is an infinitesimal constant), it means that the FL model can converge.

We analyze the proposed SSAFL under standard smoothness assumptions \cite{r41,r42} for the global objective $F(\theta) = \sum_{i=1}^{I} p_i F_i(\theta)$, where
$p_i = \frac{|D_i|}{\sum_j |D_j|}$. Recall that in each aggregation event, the server updates $\theta^{t+1} = \theta^t + \sum_{i \in Q(t)} w_i \Delta\theta_{i}^{t_i}$, where $Q(t)$ is the set of arrived clients within the micro-batch window, $t_i \le t$ is the (possibly stale) local generation time of $\Delta\theta_{i}^{t_i}$, and $w_i$ are similarity-aware aggregation weights after minimum-weight protection and renormalization. Each client $i$ uploads only if $\|\Delta\theta_{i}^{t_i}\|_2 \ge \epsilon_i$,
where $\epsilon_i = \epsilon_{\text{base}}(1+ \lambda_s(1-\text{Sim}_i(S)))$.

\subsection{Assumptions}
\begin{itemize}
\item[\textbf{A1}] (L-smoothness) Each local objective $F_i$ is $L$-smooth:
$\|\nabla F_i(\theta) - \nabla F_i(\theta')\| \le L\|\theta - \theta'\|$; hence $F$ is $L$-smooth.

\item[\textbf{A2}] (Unbiased local gradients \& bounded variance) Local stochastic gradients are unbiased
with variance $\sigma^2$: $\mathbb{E}[g_i(\theta)\,|\,\theta] = \nabla F_i(\theta)$ and
$\mathbb{E}\|g_i(\theta) - \nabla F_i(\theta)\|^2 \le \sigma^2$.

\item[\textbf{A3}] (Bounded staleness) The delay is bounded: $0 \le t - t_i \le \tau_{\max}$.

\item[\textbf{A4}] (Step sizes) Each client uses a constant stepsize $\eta \le \frac{1}{2L}$ in local SGD, with a finite number of local steps per upload.

\item[\textbf{A5}] (Weights) Aggregation weights satisfy $w_i \ge w_{\min} > 0$ for $i \in Q(t)$ and $\sum_{i \in Q(t)} w_i = 1$.

\item[\textbf{A6}] (Trigger bias control) The upload rule acts as magnitude-based sparsification: there exists $\zeta \in [0, 1)$ such that $\left\|\sum_{i \in Q(t)} \tilde{w}_i \Delta\theta_{i}^{t_i} - \sum_{i \in Q(t)} w_i \Delta\theta_{i}^{t_i}\right\| \le \zeta \left\|\sum_{i \in Q(t)} \tilde{w}_i \Delta\theta_{i}^{t_i}\right\|$,
where $\tilde{w}_i$ are the pre-weights before minimum-weight protection.
\end{itemize}

Assumption A6 is mild: with thresholded uploads and renormalization, the effective deviation from the pre-weighted update is bounded; the bound improves as $\epsilon_{\text{base}} \downarrow$ or $\lambda_s \downarrow$.

\subsection{One-step Progress}

By $L$-smoothness and the update rule, $F(\theta^{t+1}) \le F(\theta^t) + \left\langle
\nabla F(\theta^t), \sum_{i \in Q(t)} w_i \Delta\theta_{i}^{t_i} \right\rangle+ \frac{L}{2}\left\|\sum_{i \in Q(t)} w_i \Delta\theta_{i}^{t_i}\right\|^2$. Each client’s local update with step size $\eta$ and $E_i$ steps satisfies
$\mathbb{E}[\Delta\theta_{i}^{t_i}\,|\,\theta^{t_i}] \approx -\eta E_i \nabla F_i(\theta^{t_i})$ and $\mathbb{E}\|\Delta\theta_{i}^{t_i}\|^2 \le c_1 \eta^2 E_i^2(\|\nabla F_i(\theta^{t_i})\|^2 + \sigma^2)$
for some constant $c_1$ determined by the local optimizer. Using bounded staleness (A3) and smoothness, we relate stale gradients to current ones:
$\|\nabla F_i(\theta^{t_i}) - \nabla F_i(\theta^t)\| \le L\|\theta^{t_i} - \theta^t\| \le c_2 \eta \tau_{\max}$, which yields the following descent lemma.

\textit{\textbf{Lemma 1} (Descent with staleness and trigger).} Under A1--A6 and $\eta \le \frac{1}{2L}$, $\mathbb{E}\big[F(\theta^{t+1})\big] \le \mathbb{E}\big[F(\theta^t)\big]
- \eta\,\mathbb{E}\Big(1 - \frac{L\eta}{2} - \kappa_\tau - \kappa_\zeta
\Big)\,\mathbb{E}\|\nabla F(\theta^t)\|^2 + c_3 \eta^2 \Sigma$, where $E = \sum_{i \in Q(t)} w_i E_i$, $\kappa_\tau = \mathcal{O}(L\eta \tau_{\max})$ captures staleness,
$\kappa_\zeta = \mathcal{O}(\zeta)$ captures trigger/renormalization bias, and
$\Sigma = \sum_{i \in Q(t)} w_i^2 \sigma^2$.

\subsection{Main Results}

\textit{\textbf{Theorem 1} (Convex case).} If $F$ is convex and bounded below by $F^\star$, then choosing a constant $\eta \le \min\{\frac{1}{4L}, \frac{1}{2L(\tau_{\max}+1)}\}$ yields
$\frac{1}{T}\sum_{t=0}^{T-1} \mathbb{E}\|\nabla F(\theta^t)\|^2 \le
\mathcal{O}\left( \frac{F(\theta^0) - F^\star}{\eta E T}
\right) + \mathcal{O}\big(\eta\Sigma\big) + \mathcal{O}\big(L\eta \tau_{\max}\big) + \mathcal{O}(\zeta).$
In particular, with $\eta = \Theta(1/\sqrt{T})$ we obtain the standard sublinear rate
$\mathcal{O}(1/\sqrt{T})$ in terms of gradient norm, and with constant $\eta$ we get
$\mathcal{O}(1/T)$ + steady-state error governed by variance, staleness, and trigger bias.

\textit{\textbf{Theorem 2} (PL condition).} If $F$ satisfies the Polyak--\L{}ojasiewicz (PL) inequality
$\frac{1}{2}\|\nabla F(\theta)\|^2 \ge \mu(F(\theta) - F^\star)$ for some $\mu > 0$, then for
$\eta \le \min\{\frac{1}{4L}, \frac{\mu}{4L^2(\tau_{\max}+1)}\}$,
$\mathbb{E}[F(\theta^{t+1}) - F^\star] \le (1 - \mu \eta E)\,\mathbb{E}[F(\theta^t) - F^\star] + c_4\big(\eta^2 \Sigma + \eta L \tau_{\max} + \zeta\big)$, i.e., linear convergence to a neighborhood whose radius scales with variance $\Sigma$, staleness $\tau_{\max}$, and trigger bias $\zeta$.

\subsection{Remarks on Design Parameters}

\textbf{\emph{Thresholds \& similarity.}} Larger similarity $\text{Sim}_i(S)$ gives smaller $\epsilon_i$ and hence more frequent uploads; this reduces $\zeta$ (smaller trigger bias) and tightens the neighborhood in Theorem~2, at the cost of more communication. Conversely, a larger $\lambda_s$ or $\epsilon_{\text{base}}$ shrinks traffic but increases $\zeta$.

\textbf{\emph{Minimum-weight protection.}} Enforcing $w_i \ge w_{\min}$ prevents starvation of informative but low-magnitude updates, which stabilizes $E$ and improves the contraction factor $1 - \mu \eta E$.

\textbf{\emph{Staleness.}} A smaller micro-batch window and bounded network delay keep $\tau_{\max}$ small, reducing the degradation terms $\mathcal{O}(L\eta\tau_{\max})$ and improving both bounds.

Overall, SSAFL achieves standard convergence guarantees of asynchronous federated optimization under common assumptions, while its similarity-aware triggering and weighting introduce explicit, controllable trade-offs among accuracy, communication, and delay.
}

\bibliographystyle{IEEEtran}
\bibliography{main}

\vfill

\end{document}